\documentclass[12pt]{article}

\usepackage{amsmath}
\usepackage{amssymb}
\usepackage{euscript}

\usepackage{times}
\usepackage{cite}
\usepackage{amsmath}
\usepackage{amsfonts}
\usepackage{epsf}
\usepackage{epsfig}
\numberwithin{equation}{section}

\def\citobs#1{ {\bf  }~\ {\tt \bf #1} }

\def\Order#1{${\cal O}(#1)$}

\def\Oceex#1{${\cal O}(#1)_{\rm CEEX}$}
\def\KK{${\cal KK}$}

\def\st{\hbox{}} 

\def\hbeta{\hat{\beta}}

\newcommand{\Meu}{\EuScript{M}}

\newcommand{\Bmf}{\mathfrak{B}}

\newcommand{\Mmf}{\mathfrak{M}}

\newcommand{\sfac}{\mathfrak{s}}

\newcommand{\beq}{\begin{eqnarray}}
\newcommand{\eeq}{\end{eqnarray}}
\newcommand{\bea}{\begin{eqnarray}}
\newcommand{\eea}{\end{eqnarray}}
\def\ba{\begin{eqnarray}}
\def\ea{\end{eqnarray}}

\newcommand{\gamu}{\gamma_{\mu}}

\newcommand{\gafi}{\gamma_5}

\begin{document}

\thispagestyle{empty}

\begin{flushright}
{\bf  
CERN-TH/2001-161\\
DESY 01--05\\
} 
\end{flushright}
 
\noindent
\vspace*{5mm}
\begin{center}
{\LARGE\bf
     Predictions for ${\bar \nu} \nu \gamma$ Production at LEP }
\end{center}

\vspace*{2mm} 
\begin{center}
  {\bf D. Bardin$^{a}$,}~~
  {\bf S. Jadach$^{b,c,d}$,}~~
  {\bf T. Riemann$^{b}$ and}~~
  {\bf Z. W\c{a}s$^{c}$ }
  \\
  \vspace{3mm}
  {\em $^a$ Lab. of Nuclear Problems, JINR, RU-141980 Dubna, Russia}\\
  \vspace{1mm}
  {\em $^b$ Deutsches Elektronen-Synchrotron DESY,
          D-15738 Zeuthen, Germany}\\
  \vspace{1mm}
  {\em $^c$ Institute of Nuclear Physics,
           ul. Kawiory 26a, 30-055 Cracow, Poland}\\
  \vspace{1mm}
  {\em $^d$ CERN, Theory Division, 
           CH-1211 Geneva 23, Switzerland}\\
\end{center}
\vfill

\vspace*{5mm} 
\begin{abstract}
We study predictions for the reaction 
$e^+e^- \to {\bar \nu} \nu (n \gamma)$.  
The complete one-loop corrections are taken into account and  
higher order contributions, in particular those for the observed real photons, 
are added whenever necessary. 
The event generator \KK MC, a general-purpose 
Monte Carlo generator for the process $e^+e^- \to  \bar f f n\gamma$ based on the
method of exclusive exponentiation, is used as the environment.
We extend its applicability to the process  
$e^+e^- \to \bar \nu_l  \nu_l n\gamma$, $l=e,\mu,\tau$,
where the observation of at least
a single $\gamma$ is required. 
The exponentiation is implemented in much the same way
as for the $s$-channel process alone. 
In particular, all photonic effects present in the case 
of $W$ exchange, which {\it cannot}
be included in the $s$-channel exponentiation scheme,
are calculated to a finite order only.
The real hard photon matrix element is calculated  up to 
${\cal O} ({\alpha^2})$. 
Leading logarithmic contributions of the two-loop 
corrections and one-loop photonic corrections accompanying real single 
photon emission are included.
The electroweak corrections are calculated with the {\tt DIZET} 
library of the {\tt ZFITTER} package.
Numerical tests and predictions for typical observables are presented.
\end{abstract}  
\vspace*{1mm}

\begin{center}
  {\em Submitted to European Journal of Physics}
\end{center}

\vspace*{1mm}
\vfill
\begin{flushleft}
{\bf 
  CERN-TH/2001-161\\
  DESY 01--05\\
  October 2001 }
\end{flushleft}

\vspace*{1mm}
\bigskip
\footnoterule
\noindent
{\footnotesize \noindent
Work supported in part  by the European Union 5-th Framework under contract HPRN-CT-2000-00149,
NATO grant PST.CLG.977751
and
INTAS M$^{o}$ 00-00313.
\\
E-mails: bardin@nusun.jinr.ru, 
Stanislaw.Jadach@cern.ch, Tord.Riemann@desy.de, Zbigniew.Was@cern.ch  
\normalsize

\section{Introduction}
For the final LEP2 data analysis the total cross-section for  the process 
$e^+e^-\to \bar{\nu} \nu n\gamma$
will have to be calculated with a precision of $0.5\%$--$1\%$, and
arbitrary differential distributions of observable photons also
have to be calculated with similar precision~\cite{Kobel:2000aw}.
In future, for a high luminosity linear electron collider like TESLA,
the precision requirements will be even more demanding.
These requirements may be fulfilled by transferring our expertise from the 
case of charged 
lepton production to that of neutrino production, in particular the more involved case of
${\bar \nu}_e \nu_e$ with $t$-channel exchanges.
The present paper marks an important step in this direction.
It contains the necessary extensions of the Monte Carlo program 
\KK MC of ref.~\cite{Jadach:1999vf}, originally written for
$e^+e^-\to \bar{f}f,\; f=\mu,\tau,u,d,c,s,b $, to the neutrino case. 

In the neutrino pair production process 
\begin{eqnarray}
\label{mainprocess}
e^+e^- \to \bar{\nu} \nu  n\gamma,
\end{eqnarray}
one is interested
in observables where at least one high-$p_T$ photon is observed; 
neutrinos obviously escape detection. 
From a methodological point of view, however, it is 
convenient to consider 
\begin{eqnarray}
\label{starteq}
e^+e^-\to \bar{\nu} \nu 
\end{eqnarray}
as a (non-observable) Born process, and to incorporate (observable) radiative 
corrections into it,
in particular the real photon emissions, which provide the detectable signature.
A convenient method of exponentiation is discussed in this framework.
In order to achieve the 0.5\% precision level for the  $\bar{\nu} \nu \gamma$
final state,
the leading-logarithmic (LL) corrections have to be calculated up to two or three 
loops for the virtual corrections and
up to two or even three hard photons for multiple bremsstrahlung.
Mixed  real--virtual terms such as the loop corrections to 
real photon emission have to be calculated as well%
\footnote{The genuine weak one-loop corrections are sufficient.}. 
Needless to say that a sufficiently precise
integration over  the multiphoton phase space within the detector acceptance 
is also necessary.
The Monte Carlo (MC) event generator approach is the only practicable solution.

As in the case of any other two-fermion final state in $e^+e^-$ scattering, it is 
possible to define certain computational building blocks.
Our case does not require the complete
two-loop effects and we can separate the calculation into two parts:
(i) QED: interaction of photons with fermions as well as 
$WW\gamma$ and $WW\gamma\gamma$ interactions; and
(ii) the rest: 
non-photonic weak and QCD corrections.
The type (ii) corrections can be hidden in a few effective coupling constants. 

The Monte Carlo method is used for
the numerical integration over the Lorentz-invariant phase space, as usual.
The Monte Carlo event generator \KK\ MC is documented in ref.~\cite{Jadach:1999vf}.
For a detailed description of its matrix elements 
for the $e^+e^-\to \bar{f} f (n\gamma)$ processes 
we refer the reader to refs.~\cite{Jadach:1998jb,Jadach:2000ir}. 

In section 2 we discuss the implementation of the electroweak corrections.  
The package  {\tt ZFITTER}~\cite{Bardin:1999yd,home-ZFITTER} is used for this 
purpose. 
Basic numerical tests of the code in the absence of photonic effects
are described.

In  section 3 we introduce the photonic matrix elements. 
We start from the simplest cases of $\nu_\mu$ and  $\nu_\tau$ production. 
We then explain the extension of  
the matrix elements used in the CEEX exponentiation 
of refs.~\cite{Jadach:1998jb,Jadach:2000ir,Jadach:1999vf},
to the case of $e^+e^-\to  \bar{\nu_e} \nu_e\gamma$.
The modifications are due to the presence of $t$-channel $W$ exchange.
We start with the general description of our approximation,
and later present the single-photon tree-level amplitude. 
In particular, we explain  how single bremsstrahlung amplitudes 
are used as a building block in the multiple-photon amplitudes. 
Finally we briefly explain the calculation (or construction) of our amplitudes
for the different higher order cases:
one-virtual and one-real photon, and two-real photons.

In section 4, predictions of  \KK\ and KORALZ 
\cite{Jadach:1999tr,Jadach:1994yv} are given 
for selected observables of the recent LEP MC workshop~\cite{Kobel:2000aw},
including results that are used for the final estimate
of the theoretical and technical errors of our new calculation.

Section 5 concludes the paper with a statement
on the precision of our theoretical predictions for the 
${\bar \nu} \nu \gamma$ process,
as compared with the precision targets requested by LEP experiments~\cite{Kobel:2000aw}.

\section{\label{sec-effB}
The effective Born approximation for $e^+e^- \to {\bar \nu} \nu$ 
}
Similarly to the case of pure $s$-channel two-fermion processes, the 
electroweak one-loop
corrections can be incorporated via effective coupling constants of the 
$Z$ and $W$ to fermions.
Let us define here the complete electroweak one-loop corrected effective Born
cross-section for neutrino pair production%
\footnote{The notations in this 
  section follow closely those of the package {\tt ZFITTER}.}.
This is by construction a gauge-invariant quantity.
It includes $s$-channel $Z$ exchange for all three neutrino species, while
for $\nu_e$ pair production it also includes $t$-channel $W$ exchange:
\begin{eqnarray}
\label{eq-dsig}
\frac{d\sigma}{d\cos\vartheta} &=& \sum_{i=e,\mu,\tau}
\frac{d\sigma(e^+ e^- \to {\bar \nu}_i \nu_i)}{d\cos\vartheta} 
= 
~~ 3~ \sigma_s + \sigma_{st} + \sigma_t.
\label{eq-int}
\end{eqnarray}
The improved Born cross-section originates from a neutral-current matrix 
element ${\cal M}_Z$~\cite{Bardin:1992jc,Bardin:1999yd},
\begin{eqnarray}
{\cal M}_Z &=& 
\frac{G_{\mu}}{2\sqrt{2}}  \,\rho_{e\nu}^Z\,
\chi_Z(s) \,
[{\bar u}_e\gamu \left( {\bar v}_e + \gafi\right) u_e] 
\times 
[{\bar u}_{\nu} \gamu \left( 1+\gafi\right) u_{\nu}],
\label{eq-mz}
\end{eqnarray} 
and, for $ {\bar \nu}_e \nu_e $ production only, additionally from
a charged-current matrix element ${\cal M}_W$~\cite{Arbuzov:1995id}: 
\begin{eqnarray}
{\cal M}_W &=&
 \frac{G_{\mu}}{\sqrt{2}} \,\rho^W_{e{\nu}_e}\,
\chi_W(t)\, 
[{\bar u}_e \gamu \left( 1+\gafi\right) u_e] 
\times 
[{\bar u}_{\nu} \gamu \left( 1+\gafi\right) u_{\nu}].
\label{eq-mw}
\end{eqnarray} 
We use here the notations $a_e = a_{\nu} =1$, $Q_e = -1$, 
$s_W^2 = 1 - M_W^2 / M_Z^2$, and have only three form factors $\kappa_e$,
$\rho_{e\nu}^Z$, and $\rho^W_{e{\nu}_e}$:
\begin{eqnarray}
{\bar{v}}_{e}  &=& 
1 - 4 |Q_e | s_W^2 \, \kappa_e.
\label{labve} 
\end{eqnarray}
In the Born approximation, it is $\rho = \kappa = 1$.
The kinematical invariants are used in the approximation
$m_e=0$: 
\begin{eqnarray}
  \label{def-t}
  t &=& - \frac{s}{2} (1-\cos\vartheta),\\
  u = -s-t  &=& - \frac{s}{2} (1+\cos\vartheta).
\label{def-u}
\end{eqnarray}
We also use:
\begin{eqnarray}
  \chi_B(s) &=& 
  \frac{M_B^2}{-s+M_B^2(s)},
  \\
  \label{mass}
  M_B^2(s) &=& M_B^2 -i M_B ~\Gamma_B(s) ~\theta(s),
  \label{chi}
\end{eqnarray}
and the width in the $s$-channel may be chosen constant or $s$-dependent.
The resulting cross-section contributions are:
\begin{eqnarray}
  \label{eq-sigs}
  \sigma_s &=& 
  \frac{s G_{\mu}^2}{128\pi} 
  \left| \chi_Z(s) ~ \rho^Z_{e\nu}\right|^2 
  \Bigl[   
  (1+\cos^2\vartheta)  ~  (1 + |v_{e}|^2)
  + 4~\cos\vartheta  ~ ~\Re e ~v_{e}
  \Bigr] ,
  \\
  \sigma_{st} &=& 
  -   \frac{s G_{\mu}^2}{32\pi} 
  ~\Re e \Bigl\{
  \chi_Z(s) \chi_W^*(t)  ~ \rho^Z_{e\nu} ~ \rho^{W*}_{e\nu}
  (1+\cos\vartheta)^2 ~  (1+v_{e})
  \Bigr\},
  \label{eq-sigst}
  \\
  \sigma_{t} &=& 
  \frac{s G_{\mu}^2}{16\pi} 
  \left| \chi_W(t) \rho^W_{e\nu}\right|^2 
  (1+\cos\vartheta)^2.
  \label{eq-sigt}
\end{eqnarray}

The weak neutral form factors $\kappa_e$
and $\rho_{e\nu}^Z$ are discussed in detail in 
\cite{Bardin:1989di,Bardin:1999yd} and in references therein%
\footnote{%
  These form factors are calculated in the library {\tt DIZET} as variables
  {\tt XROK(2)} and  {\tt XROK(1)} by calling
  subroutine {\tt ROKANC(u,-s,t)} for $s>0$ and $t,u<0$.}. 
For latest comparisons and applications at LEP1, see also
\cite{Bardin:1995aa}, and at LEP2, see ref.~\cite{Kobel:2000aw}.
There is one modification with respect to earlier applications, which has 
to be clarified here.
Even though the complete virtual corrections are not infrared-finite 
(because of photonic diagrams), we prefer {\em not} to split 
away the photonic part of the virtual corrections in our formulae. 
Instead, we prefer to  
combine the complete virtual corrections with real bremsstrahlung from the 
package%
\footnote{%
  Of course $\cal KK$ MC includes virtual QED corrections of its own. 
  The subtraction of the QED part from the complete loop corrections 
  will be defined in  section 2.1.}
\KK\ MC.
For this reason, we calculate with the weak library {\tt DIZET} 
of the package {\tt ZFITTER} the complete virtual correction $\rho_{e\nu}^Z$.
Since in the quantity {\tt  XROK(1)} the QED part is explicitly subtracted, 
we have to re-establish its contribution here
according to the formula: 
\begin{eqnarray}
  \label{masterrhoz}
  \rho_{e\nu}^Z &=& 
  {\tt XROK}(1) + {\tt QED\_NC},
  \\
  \label{mastrhoz}
  {\tt QED\_NC} &=&
  \frac{\alpha}{2\pi} Q_e^2 
  \left[ -\left( L_e -1 \right)\ln\frac{m_e^2}{\lambda^2}
    -\frac{1}{2} L_e^2
    +\frac{3}{2} L_e
    +4~ {\rm Li}_2(1)-2
  \right],
\end{eqnarray}
with $L_e=\ln (s/m_e^2)$ and $\lambda$ a finite photon mass.

The charged current form factor may be extracted from derivations done for 
$ep$ scattering~\cite{Bardin:1982sv,Bardin:1987rz,Bardin:1989vz}. 
Again, an infrared-finite quantity {\tt ROWB}%
\footnote{%
  With a call on subroutine {\tt RHOCC(u,-t,s)} (for $s>0$ and $t,u<0$).
  This infrared-finite term was determined by a subtraction of 
  (gauge-dependent) photonic corrections, which were combined for applications 
  at HERA with real-photon corrections.  
  Note also that we use for this part the names of subroutines (but not 
  of variables) from the package {\tt HECTOR}\cite{Arbuzov:1995id}.
  }
was constructed, rather than the full form factor $ \rho_{e\nu}^W $:
\begin{eqnarray}
  \label{mastrhow}
  \rho_{e\nu}^W &=& 
  {\tt ROWB} + {\tt QED\_CC},
  \\
  {\tt QED\_CC} &=&
  \frac{\alpha}{2\pi} Q_e^2  
  \left[
    - \left( L_e-1\right)
    \ln\frac{m_e^2}{\lambda^2} 
    -\frac{1}{2}L_e^2
    +\frac{3}{2} \ln \frac{M_W^2}{m_e^2}
    + \frac{1}{2} \ln^2\frac{t}{s}  
  \right].
\end{eqnarray}
For practical reasons, we have decided
to insert the neutral current term {\tt QED\_NC} 
(instead of {\tt QED\_CC}) also in the charged current case:
\begin{eqnarray}
  \label{masterrhow}
  \rho_{e\nu}^{W} &=& 
  {\tt ROWB'} + {\tt QED\_NC},
  \\
  \label{addit3}
  {\tt ROWB'} &=& {\tt ROWB}+ {\tt QED\_CC} - {\tt QED\_NC}.
\end{eqnarray}
All the virtual corrections described here come with {\tt ZFITTER} v.6.36
(21 June 2001)~\cite{Bardin:1999yd,home-ZFITTER}.

\subsection{Form factors and \KK\ }
The effects due to the loop diagrams given in formulae (\ref{masterrhoz}) 
and  (\ref{masterrhow})
are separated into the finite parts, which are encapsulated in two of the 
electroweak form factors and pretabulated in \KK\ in order to save computer
time, and the infrared-divergent part {\tt QED\_NC} 
which defines the (now universal) genuine QED corrections. 
The term {\tt QED\_NC} is not included in the form factors, but enter the 
QED part of the calculation (encapsulated in  \KK\ MC).
The remaining finite parts modify $Z$ 
and $W$ couplings to fermions as in Refs.~\cite{Jadach:1998jb,Jadach:2000ir}.

We are currently not aiming yet at a calculation of the complete second-order corrections. 
We may, therefore, incorporate the difference between the QED parts for $s$- and $t$-channel
\begin{equation}
  \label{delta_CC-NC}
  \delta_{\tt CC-NC}={\tt QED\_CC} - {\tt QED\_NC}= \frac{\alpha}{2\pi} Q_e^2 
  \left[
    \frac{3}{2} \ln \frac{M_W^2}{s} + \frac{1}{2} \ln^2\frac{t}{s} 
    -4~ {\rm Li}_2(1)+2
  \right]
\end{equation}
into the electroweak formfactor ${\tt ROWB'}$, as mentioned above.
The difference (\ref{delta_CC-NC})
is infrared-finite, numerically small, and not enhanced by the
large logarithms with respect to $\frac{\alpha}{2\pi}$. 
The numerical contribution is in fact below 0.5\% of the Born cross section
for LEP2 energies, when integrated over neutrino angular variables.

\subsection{Technical tests for the ``academic'' event selection}

\begin{table}[!ht]
\small
\centering
\caption{\small
Electroweak effects including pretabulation as implemented in \KK\ MC.
Total cross--sections and forward-backward asymmetries are calculated without 
QED corrections.
For every total energy the upper entry is from \KK sem and the lower one
from {\tt ZFITTER}.
}
\begin{tabular} {||l|l|l|l||}
\hline\hline
Channel                            &
Cms energy   [GeV]                 &
$\sigma $                          &
$A_{FB} $                                              
\\
\hline\hline
$e^+e^- \to \mu^+\mu^-$     & 
      91.19  &  2.00345818887006  &  0.01782302636524   \\ 
  &          &  2.00339325242333  &  0.01783385260585   \\
\hline
  &  100.00  &  0.05261755867120  &  0.58632495479977   \\    
  &          &  0.05261818913548  &  0.58632716129114   \\
\hline
  &  140.00  &  0.00697974693325  &  0.66477236253951   \\    
  &          &  0.00697977362831  &  0.66477185420337   \\
\hline
  &  189.00  &  0.00337662390496  &  0.56552465469245   \\    
  &          &  0.00337666453508  &  0.56552686535458   \\
\hline
  &  200.00  &  0.00298389425320  &  0.55492797790038   \\    
  &          &  0.00298394853103  &  0.55493448086923   \\
\hline
  &  206.00  &  0.00279941156110  &  0.54984466446977   \\    
  &          &  0.00279947462003  &  0.54985275218873   \\                               
\hline
\hline\hline
$e^+e^- \to \bar\nu_\mu\nu_\mu$    &
      91.19  &  3.97432928812849 &   0.10951545488039   \\
   &         &  3.97416166784039 &   0.10954931699922   \\
\hline
   & 100.00  &  0.08476335609986 &   0.10779125894503   \\
   &         &  0.08476366520697 &   0.10782386820347   \\ 
\hline
   & 140.00  &  0.00375688589078 &   0.10221239752831   \\
   &         &  0.00375689763895 &   0.10224383875601   \\
\hline
   & 189.00  &  0.00115026771057 &   0.09199984287991   \\
   &         &  0.00115029626002 &   0.09203384569328   \\ 
\hline
   & 200.00  &  0.00096201292991 &   0.08639737806323   \\
   &         &  0.00096202811147 &   0.08642343541349   \\
\hline
   & 206.00  &  0.00087902320002 &   0.08379052580837   \\
   &         &  0.00087903346591 &   0.08381353397919   \\
\hline
\hline\hline
$e^+e^- \to \bar \nu_e\nu_e$       &
      91.19  &  3.98484724572530  &  0.11158999944974   \\
  &          &  3.98468359582307  &  0.11162475358080   \\
\hline  
  &  100.00  &  0.15558776383424  &  0.44218368752900   \\
  &          &  0.15560448223749  &  0.44225943946968   \\ 
\hline  
  &  140.00  &  0.04084986326864  &  0.82963857483597   \\
  &          &  0.04086399163970  &  0.82969877450851   \\ 
\hline  
  &  189.00  &  0.03937265031133  &  0.91661747898646   \\ 
  &          &  0.03939257132943  &  0.91665886557827   \\ 
\hline 
  &  200.00  &  0.03971937632138  &  0.92595698881602   \\
  &          &  0.03974060496062  &  0.92599592232744   \\
\hline  
  &  206.00  &  0.03993258937665  &  0.93040498709775   \\  
  &          &  0.03995453232877  &  0.93044271454420   \\ 
\hline
\hline\hline
\end{tabular}
\label{born-tab}
\end{table}

Before the actual discussion of the bremsstrahlung part of the 
generator, let us make certain elementary numerical tests
of the implementation of the $\rho_{e\nu}^{W}$ function
of eq.~(\ref{masterrhow}) within the \KK MC program.
This will include the neutral current form factors as well, 
and will be done with the aid of the semi-analytical
package {\KK}sem, the internal testing program of {\KK} MC.
As a first step we compare the effective Born predictions
as calculated by  {\KK}sem and by {\tt ZFITTER}.
In table \ref{born-tab} we present the  predictions from the two programs 
for the processes $e^+e^- \to \mu^+ \mu^-$,
$e^+e^- \to \bar \nu_\mu \nu_\mu$, 
and finally for the processes $e^+e^- \to \bar \nu_e \nu_e$. 
In all cases the agreement between {\KK}sem and {\tt ZFITTER}
we find to be sufficiently good,
slightly less precise in case of $e^+e^- \to \bar \nu_e \nu_e$. 
In the latter case possible uncertainties of numerical integration
(and pretabulation) can be the reason, as the distributions of the scattering angle
peakes in the forward region.
We conclude that the program \KK\ MC
properly exploits the electroweak form factors of the package {\tt DIZET}.
This opens the way to a more complete treatment of the EW corrections
in \KK\ MC for the neutrino channels.

\section{Exponentiation and $t$-channel $W$-exchange}

The coherent exclusive exponentiation (CEEX) was introduced  
in~refs.~\cite{Jadach:1998jb,Jadach:2000ir}.
It is deeply rooted in the Yennie Frautschi Suura (YFS) exponentiation~\cite{yfs:1961},
and for applications to  narrow resonances see also
earlier related refs.~\cite{Greco:1975ke,greco:1980}.
The exponentiation procedure, i.e. the re-organization of the QED perturbative 
series such that infrared (IR) divergences are summed up to infinite order,
is done for both real and virtual emissions
at the spin-amplitude level (the case of CEEX), 
while the actual cancellation of the IR divergences
always occurs at the integrated cross-section level.
CEEX is an extension of the traditional YFS exponentiation, in the sense that,
in  the standard YFS exponentiation 
(which we call EEX -- for exclusive exponentiation),
the isolation of the real IR divergences is done
after squaring and spin-summing spin amplitudes,
while in CEEX it is done before.
In the actual implementation of the CEEX,
all spin amplitudes for the fermion pair production
in electron--positron scattering are handled
with help of the powerful Weyl spinor (WS) techniques.
There are several variants of the WS techniques.
We have chosen the method of Kleiss and Stirling 
(KS)~\cite{kleiss-stirling:1985,Kleiss:1986qc},
because we found KS method well suited for constructing the multiphoton spin amplitudes.
In refs.~\cite{Jadach:1998jb,Jadach:2000ir} more details of the approach are available.
In particular we take all notations and definitions from these works. 
In the following we recall
only the very basic formulae of refs.\cite{Jadach:1998jb,Jadach:2000ir},
before we show how the multiphoton  CEEX spin amplitudes for the
$W$ contribution $t$-channel is constructed and calculated.

\subsection{The master formula}
Defining the Lorentz-invariant phase space as
\begin{equation}
\label{eq:lips}
\int d{\rm Lips}_n(P;p_1,p_2,...,p_n) 
    = \int (2\pi)^4\delta(P-\sum_{i=1}^n p_i) 
\prod_{i=1}^n \frac{ d^3 p}{(2\pi)^3 2p^0_i},
\end{equation}
we write the general CEEX total cross-section for the process
\begin{equation}
e^+(p_a) +e^-(p_b) 
  \to f(p_c) +\bar{f}(p_d) +\gamma(k_1) +\gamma(k_2)+...+\gamma(k_n), n=0,1,2,...,\infty,
\end{equation}
with polarized beams and decays of unstable final fermions
being sensitive to fermion spin polarizations
(neutrinos are, of course, taken as stable),
as follows~\cite{Jadach:1998jb}:
\begin{equation}
  \label{eq:sigma-ceex2}
  \sigma^{(r)} =  {1\over {\rm flux}(s)}
  \sum_{n=0}^\infty 
  \int d{\rm Lips}_{n+2} ( p_a+p_b; p_c,p_d, k_1,\dots,k_n)\;
  \rho^{(r)}_{\rm CEEX}  ( p_a,p_b, p_c,p_d, k_1,\dots,k_n)
\end{equation}
where
\begin{equation}
  \label{eq:rho-ceex2}
  \begin{split}
  \rho^{(r)}_{\rm CEEX} &( p_a,p_b, p_c,p_d, k_1,k_2,\dots,k_n)=
  {1\over n!} e^{Y(\Omega;p_a,...,p_d)}\;\bar{\Theta}(\Omega)\;
    \sum_{\sigma_i=\pm 1}\;
    \sum_{\lambda_i,\bar{\lambda}_i=\pm 1}\;
\\&
    \sum_{i,j,l,m=0}^3\;
        \hat{\varepsilon}^i_a                  \hat{\varepsilon}^j_b\;
        \sigma^i_{\lambda_a \bar{\lambda}_a}   \sigma^j_{\lambda_b \bar{\lambda}_b}
    \Mmf^{(r)}_n 
    \left(\st^{p}_{\lambda} \st^{k_1}_{\sigma_1} \st^{k_2}_{\sigma_2}
                                           \dots \st^{k_n}_{\sigma_n} \right)
    \left[
    \Mmf^{(r)}_n 
    \left(\st^{p}_{\bar{\lambda}} \st^{k_1}_{\sigma_1} \st^{k_2}_{\sigma_2}
                                                 \dots \st^{k_n}_{\sigma_n} \right)
    \right]^\star
        \sigma^l_{\bar{\lambda}_c \lambda_c }   \sigma^m_{\bar{\lambda}_d \lambda_d }
        \hat{h}^l_c                             \hat{h}^m_d.
  \end{split}
\end{equation}
Assuming the dominance of the $s$-channel exchanges, including resonances,
we define the complete set of spin amplitudes
for the emission of $n$ photons in \Oceex{\alpha^r} ($r=0,1,2$) 
as follows:
\begin{equation}
\begin{split}
  \label{eq:ceex-master}
  &\Mmf^{(0)}_n\left( \st^{p}_{\lambda} \st^{k_1}_{\sigma_1}
                                   \dots \st^{k_n}_{\sigma_n}  \right)
   =\!\!\!\sum_{\wp\in\{I,F\}^n}\;  
     \prod_{i=1}^n \; \sfac^{\{\wp_i\}}_{[i]}\;
      \beta^{(0)}_0 \left( \st^{p}_{\lambda}; X_\wp \right),
\\
  &\Mmf^{(1)}_n\left( \st^{p}_{\lambda} \st^{k_1}_{\sigma_1}
                                   \dots \st^{k_n}_{\sigma_n}  \right)
   =\!\!\!\sum_{\wp\in\{I,F\}^n}\;  
     \prod_{i=1}^n \; \sfac^{\{\wp_i\}}_{[i]}\;
     \left\{ \beta^{(1)}_0 \left( \st^{p}_{\lambda}; X_\wp \right)
            +\sum_{j=1}^n 
             {\beta^{(1)}_{1\{\wp_j\}} \left( \st^{p}_{\lambda} \st^{k_j}_{\sigma_j} ; X_\wp \right)
                                          \over \sfac^{\{\wp_j\}}_{[j]} }\;
    \right\},
\\ 
  &\Mmf^{(2)}_n\left( \st^{p}_{\lambda} \st^{k_1}_{\sigma_1}
                                   \dots \st^{k_n}_{\sigma_n}  \right)
   =
\\
   &\!\!\!\sum\limits_{\wp\in\{I,F\}^n}\;  
    \prod_{i=1}^n \; \sfac^{\{\wp_i\}}_{[i]}\;
    \left\{  \beta^{(2)}_0 \left( \st^{p}_{\lambda}; X_\wp \right)
            +\sum_{j=1}^n 
             {\beta^{(2)}_{1\{\wp_j\}} \left( \st^{p}_{\lambda} \st^{k_j}_{\sigma_j} ; X_\wp \right)
                                          \over \sfac^{\{\wp_j\}}_{[j]} }\; 
             +\sum\limits_{1\leq j<l\leq n}\;\!\!\!\!
              {\beta^{(2)}_{2\{\wp_j\wp_l\}}
                   \left( \st^{p}_{\lambda} \st^{k_j}_{\sigma_j}  \st^{k_l}_{\sigma_l} ;X_\wp \right)
                               \over \sfac^{\{\wp_j\}}_{[j]} \sfac^{(\wp_l)}_{[l]} }\;
    \right\}.
\end{split}
\end{equation}
In the following, we will explain only those aspects of the above formulae
that are new for the $t$-channel $W$-boson implementations (e.g. the phase space
and general form of terms will remain untouched);
for all the rest we refer the reader to refs.~\cite{Jadach:1998jb,Jadach:2000ir}. 
We will start from the simplest case and consecutively
explain how to add more complicated terms. 

Obviously, since only initial-state bremsstrahlung contributes, sums over the 
partitions drops out.
The emissions from $W$ bosons (if present) will not be treated  as an additional
source of emission requiring a special partition for itself. 
It will play the role of
a correction to the (infrared-finite) $\beta$-functions at any perturbative order.

\subsection{ Case of $\nu_\mu$ and $\nu_\tau$ production}
For all neutrino flavours except $\nu_e$, we can limit ourselves
to changes of the numerical values of the $Z$ couplings from  quarks or leptons
to  neutrinos. 
All the formulae remain unchanged otherwise, and the previously estimated
precisions for the muon channel remain valid also for the neutrino case.
The only new element is that of the implementation of electroweak form 
factors for the neutrinos.
They were not investigated for the neutrino channel until the present paper,
for DIZET being either a weak library of \KK\ MC or a stand-alone code.

\subsection{The $\nu_e$ implementation, general aims}

In the case of calculations performed at any fixed order, there 
is, in principle, no need 
to worry about gauge invariance, cancellation
of infrared or ultraviolet singularities, etc. 
However, in practice it is sometimes quite non-trivial to 
achieve these essentials.
Already the introduction of the $W$ and $Z$ boson propagators require summation 
of infinite series of partial contributions from any order of the perturbation 
expansion.
Otherwise, the cross-sections at the peak of 
e.g. the $Z$ resonance would not be well defined. 
There are standard techniques,
such as the renormalization group, structure functions,
and exponentiation, for summing up leading higher-order terms arising from ultraviolet,
infrared, or collinear singularities.

Our ambitions here are rather limited.  We want to exploit the
relative similarity of the spin amplitudes involving $t$-channel
$W$ exchange and $s$-channel $Z$ exchange at relatively low energies.
Even though diagrams for electron neutrino pair production involving
$W$ exchange with photon lines attached only to electron lines are not
gauge-invariant (contrasting with the case with the analogous $s$-channel $Z$
exchange), the necessary contribution for completing the gauge-invariant
amplitude is small (and even vanishes within the LL approximation).  
$W$-exchange also drops out at sufficiently low energies, where it is legitimately
approximated as a contact interaction.
As a first step,
we shall exploit the simplified amplitudes for the $W$-exchange in the contact interaction
approximation, which formally will look
exactly the same as the contribution of an additional heavy $Z'$.
As a second step we shall calculate the
difference of the correct/complete perturbative results at fixed,
first and second order, with the above approximation.
In such a two-step procedure we can easily use the already
developed and tested formulation for the $s$-channel CEEX
and the corresponding program code.
In addition, we shall also be capable
dividing the $W$-exchange amplitude into individually gauge-invariant
parts of a well-defined physical origin.

In the following we will explain our approach in more detail. 
We start with a common technical trick exploiting Fierz transformation:
for massless neutrinos it is possible to rearrange the 
lowest order $e^+e^- \to \nu \bar \nu $ $W$-exchange amplitude  into  a
form identical to the one for $Z'$ production exchange,
except that the propagator now depends on $t$ and  the coupling constants
are redefined.
The complete Born-level spin amplitude then reads
\begin{equation}
  \label{eq:born-def}
  \begin{split}
     &\Bmf\left(\st^{p}_{\lambda}; X \right)=
      \Bmf\left(  \st^{p_a}_{\lambda_a} \st^{p_b}_{\lambda_b}
                  \st^{p_c}_{\lambda_c} \st^{p_d}_{\lambda_d}; X \right)=
      \Bmf\left[  \st^{p_b}_{\lambda_b} \st^{p_a}_{\lambda_a}\right]
          \left[  \st^{p_c}_{\lambda_c} \st^{p_d}_{\lambda_d}\right]\!(X)=
      \Bmf_{[bc][cd]}(X)=
\\  &\qquad\qquad
      =ie^2 \sum_{B=\gamma,Z,W} \Pi^{\mu\nu}_B(X)\; (G^{B}_{e,\mu})_{[ba]}\; (G^{B}_{f,\nu})_{[cd]}\; H_B
      =\sum_{B=\gamma,Z,W} \Bmf^B_{[bc][cd]}(X),
\\
     &(G^{B}_{e,\mu})_{[ba]} \equiv \bar{v}(p_b,\lambda_b) G^{B}_{e,\mu} u(p_a,\lambda_a),\;\;
      (G^{B}_{f,\mu})_{[cd]} \equiv \bar{u}(p_c,\lambda_c) G^{B}_{f,\mu} v(p_d,\lambda_d),
\\
     &G^{B}_{e,\mu} = \gamma_\mu \sum_{\lambda=\pm} \omega_\lambda g^{B,e}_\lambda,\quad
      G^{B}_{f,\mu} = \gamma_\mu \sum_{\lambda=\pm} \omega_\lambda g^{B,f}_\lambda,\quad
      \omega_\lambda = {1\over 2}(1+\lambda\gamma_5),
\\
     &\Pi^{\mu\nu}_{B=Z,\gamma}(X) = { g^{\mu\nu} \over X^2 - {M_{B}}^2 +i\Gamma_{B} {X^2 / M_{B}} },
\\
     &\Pi^{\mu\nu}_{B=W}(X) = { g^{\mu\nu} \over t_0 - {M_{W}}^2  }.
  \end{split}
\end{equation}
Similarly to the case of $Z$ and $\gamma$ exchanges, described
in ref.~\cite{Jadach:2000ir}, the $W$ contribution takes the following form:
\begin{equation}
  \label{born}
    \Bmf^W_{[ba][cd]}(X)
 = 2 ie^2 \frac{ 
            \delta_{\lambda_a, -\lambda_b}
            \big[\;
                 g^{W,e}_{ \lambda_a} g^{W,f}_{-\lambda_a}\;
                 T_{ \lambda_c \lambda_a}\; T'_{\lambda_b \lambda_d}
                +g^{W,e}_{ \lambda_a} g^{W,f}_{ \lambda_a}\;
                 U'_{ \lambda_c \lambda_b}\; U_{\lambda_a \lambda_d}
            \big]
         }
         { t - {M_{W}}^2  },
\end{equation}
where $t$ is calculated uniquely for  every individual  event from the
$4$-momenta of%
\footnote{Except some contributions 
  to $\beta_1$ $\beta_2$, where the ``correct'' transfers involving momenta of the 
  photons has to be re-established.
  See discussion below.}
$e^+,e^-,\nu,\bar\nu$,
$g^{W,e}_{\lambda=-1,1}=- {1 \over \sqrt{2} \sin\theta_W}, 0$ and  
$g^{W,\nu_e}_{\lambda=-1,1}=0,{1 \over \sqrt{2} \sin\theta_W}$.

Let us now consider the amplitudes involving the emission of real photons.
We also apply the Fierz transformation. 
The contribution $ \Bmf^W$ is then  added 
at {\it any}  place where the standard $Z$ contribution (eq. (44) 
of ref.~\cite{Jadach:2000ir}) occurs --
that is, in the definition of all $\beta_i^{(j)}$.
Note that such an approximation can be used at any order of the perturbation 
expansion.
This approximation preserves gauge invariance, because the resulting spin amplitudes
look formally as a contribution from an additional heavy
$Z'$, with the appropriately chosen coupling constants and propagators.

The recipe for extending the \KK\ MC amplitudes to the $\nu_e$-pair production 
at this introductory level is relatively simple:
modify eq.~(43) of ref.~\cite{Jadach:2000ir}
and use it later on as a building block for all other amplitudes,
virtual corrections and hard bremsstrahlung alike%
\footnote{In the program this is realized 
  by calling subroutine {\tt  GPS$\_$BornWPlus} from subroutine 
  {\tt GPS$\_$BornPlus}.}.
For the processes at low energies (substantially lower than 80 GeV
CMS energy) such an approximation coincides with the standard
approximation  of the contact interaction, where the momentum transfer is 
completely neglected in the $W$ propagators and the $W$--$\gamma$ vertices are neglected. 

At higher energies, we have to take the $t$-channel transfer into account; 
this modifies the structure of the amplitudes significantly.
We may still, as an intermediate solution, introduce an auxiliary single ``mean'' 
(``effective'') transfer $t_0$ for the entire amplitude
(including propagators in loop diagrams).
With such an auxiliary transfer,
the structure of the whole set of spin
amplitudes still coincides with the one for $s$-channel processes,
provided that we also drop all $W$--$\gamma$ interactions.
The introduction of such an auxiliary transfer $t_0$ is the source of
certain ambiguity in the case of an event with hard photons. 
The optimal choice of $t_0$ should, of course, minimize
the unaccounted higher-order corrections.

At present, the routine {\tt KinLib$\_$ThetaD} of \KK\ MC is used 
to define the transfer $t_0$, the same as for the
calculation of the $\theta$-dependent box corrections 
in the earlier published versions of the \KK\ MC.

\subsubsection{One real photon}
The starting point is the well-known \Order{\alpha} spin amplitude for
the single-photon brems\-strahlung. We have to consider it anew, because
we need it in conventions of  ref.~\cite{Jadach:2000ir}.
In particular we need to  keep track of the relative complex phases of parts
of the  amplitude, which enter the soft photon factors and the remaining
finite parts. 
Also, we want to (re)use the part of the amplitude
for $\nu_\mu$ channel in the $\nu_e$ case without any modifications.
This will be a starting point for obtaining $\hbeta^{(0)}_1$
of eq.~(\ref{eq:ceex-master}), which will be later incorporated into our 
general scheme of exponentiation,
exactly as explained in~\cite{Jadach:2000ir}.
\begin{figure}[h]
\centering
\setlength{\unitlength}{1mm}
\begin{picture}(180,85)
\put( -75, -35){\makebox(0,0)[lb]
{\epsfig{file=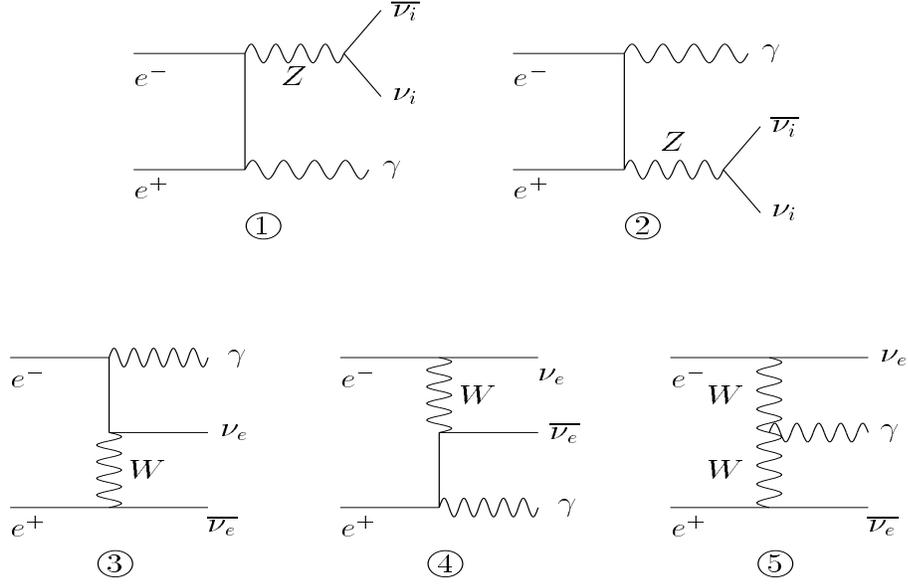,width=280mm,height=350mm}}}
\end{picture} 
 \caption{{\em The Feynman diagrams for
          $e^+ e^- \rightarrow \bar \nu_e \nu_e \gamma$.}}
 \label{fig:bremI}
\end{figure}

The first-order matrix element from the Feynman diagrams
depicted in fig.~\ref{fig:bremI}  reads
\begin{equation}
  \label{isr-feynman}
  \begin{split}
    \Meu_{1\{I\}}\left( \st^{p}_{\lambda} \st^{k_1}_{\sigma_1} \right)=
   & eQ_e \;
    \bar{v}(p_b,\lambda_b)\; \mathbf{M}^{bd}_{\{I\}}\;
         {\not\!{p_a}+m-{\not\!k_1} \over -2k_1p_a} \not\!{\epsilon}^\star_{\sigma_1}(k_1)\;
    u(p_a,\lambda_a)\\
   +&eQ_e \;
    \bar{v}(p_b,\lambda_b)
         \not\!{\epsilon}^\star_{\sigma_1}(k_1)\; {-{\not\!p_b}+m+\not\!{k_1} \over -2k_1p_b} \;
         \mathbf{M}^{ac}_{\{I\}}\;
    u(p_a,\lambda_a) \\
    &+e \;
    \bar{v}(p_b,\lambda_b)
        \; \;
         \mathbf{M}^{bd,ac}_{\{I\}}\;
    u(p_a,\lambda_a) {\epsilon}^\star_{\sigma_1}(k_1) \cdot (p_c-p_a+p_b-p_d)
\\      \;
   +&e \; \bar{v}(p_b,\lambda_b)  g_{\lambda_b,\lambda_d}^{We\nu}
         \not\!{\epsilon}^\star_{\sigma_1}(k_1)\;  \;
    v(p_d,\lambda_d)
    \bar{u}(p_c,\lambda_c)  g_{\lambda_c,\lambda_a}^{We\nu}
         \not\!k_1\;  \;
    u(p_a,\lambda_a)\\
   -&e \; \bar{v}(p_b,\lambda_b)  g_{\lambda_b,\lambda_d}^{We\nu}
         \not\!k_1\;  \;
    v(p_d,\lambda_d)
    \bar{u}(p_c,\lambda_c) g_{\lambda_c,\lambda_a}^{We\nu}
         \not\!{\epsilon}^\star_{\sigma_1}(k_1)\;  \;
    u(p_a,\lambda_a),
  \end{split}
\end{equation}
or, equivalently, as follows:
\begin{equation}
  \label{isr-feynman1}
  \begin{split}
   &\Meu_{1\{I\}}\left( \st^{p}_{\lambda} \st^{k_1}_{\sigma_1} \right)=
    {\cal M}^0+{\cal M}^1_{WW\gamma}+{\cal M}^2_{WW\gamma}+{\cal M}^3_{WW\gamma}
   \\&
    {\cal M}^0=
   eQ_e \;
    \bar{v}(p_b,\lambda_b)\; \mathbf{M}^{bd}_{\{I\}}\;
         {\not\!{p_a}+m-{\not\!k_1} \over -2k_1p_a} \not\!{\epsilon}^\star_{\sigma_1}(k_1)\;
    u(p_a,\lambda_a)
    \\&
   +eQ_e \;
    \bar{v}(p_b,\lambda_b)
         \not\!{\epsilon}^\star_{\sigma_1}(k_1)\; {-{\not\!p_b}+m+\not\!{k_1} \over -2k_1p_b} \;
         \mathbf{M}^{ac}_{\{I\}}\;
    u(p_a,\lambda_a) 
    \\&
       {\cal M}^1=+e \;
    \bar{v}(p_b,\lambda_b)
        \; \;
         \mathbf{M}^{bd,ac}_{\{I\}}\;
    u(p_a,\lambda_a) {\epsilon}^\star_{\sigma_1}(k_1) \cdot (p_c-p_a+p_b-p_d) {1 \over {t_a -M_W^2}}{1 \over {t_b -M_W^2}},
    \\&
    {\cal M}^2=   +e \; \bar{v}(p_b,\lambda_b)  g_{\lambda_b,\lambda_d}^{We\nu}
         \not\!{\epsilon}^\star_{\sigma_1}(k_1)\;  \;
    v(p_d,\lambda_d)
    \bar{u}(p_c,\lambda_c)  g_{\lambda_c,\lambda_a}^{We\nu}
         \not\!k_1\;  \;
    u(p_a,\lambda_a) {1 \over {t_a -M_W^2}}{1 \over {t_b -M_W^2}}
    \\&
     {\cal M}^3=  -e \; \bar{v}(p_b,\lambda_b)  g_{\lambda_b,\lambda_d}^{We\nu}
         \not\!k_1\;  \;
    v(p_d,\lambda_d)
    \bar{u}(p_c,\lambda_c) g_{\lambda_c,\lambda_a}^{We\nu}
         \not\!{\epsilon}^\star_{\sigma_1}(k_1)\;  \;
    u(p_a,\lambda_a) {1 \over {t_a -M_W^2}}{1 \over {t_b -M_W^2}},
  \end{split}
\end{equation}
where 
\begin{equation}
 \mathbf{M}^{xy}_{\{I\}} =ie^2 \sum_{B = W,Z} \Pi^{\mu\nu}_B(X)\; 
                              G^{B}_{e,\mu}\; (G^{B}_{f,\nu})_{[cd]}
\end{equation}
is the annihilation scattering spinor matrix, including final-state spinors and
\begin{equation}
 g_{\lambda_c,\lambda_a}^{We\nu} =e {1 \over \sqrt{2} \sin\theta_W} 
\delta^{\lambda_c}_{\lambda_a} \delta^{\lambda_c}_+.
\end{equation}
For the $W$ contribution, the subscripts in $ \mathbf{M}_{\{I\}}$ 
define the momentum
transfer in the $W$ propagator $\Pi^{\mu\nu}_W(X)$: for $ac$
the transfer is $t_a=(p_a -p_c)^2$, for $bd$ it is  $t_b=(p_b -p_d)^2$. 
If  both are explicitly marked, then the amplitude
\begin{equation}
  \mathbf{M}^{bd,ac}_{\{I\}} =ie^2 
         (G^{W}_{e,\mu})_{[ba]}\; (G^{W,\; \mu}_{\nu})_{[cd]}
\end{equation}
has to be used. 

We split the above expression into soft IR parts proportional to $(\not\!{p} \pm m)$
and non-IR parts proportional to ${\not\!k_1}$.
Employing the completeness relations of eq.~(A14) from ref.\cite{Jadach:2000ir}
to those parts we obtain:
\begin{equation}
  \begin{split}
    \Meu_{1\{I\}}
    \left( \st^{p}_{\lambda} \st^{k_1}_{\sigma_1}
    \right)=
   &-{eQ_e\over 2k_1p_a}\; \sum_\rho
     \Bmf\left[ \st^{p_b}_{\lambda_b}  \st^{p_a}_{\rho_a}\right]\st_{[cd]}
          U\left[ \st^{p_a}_{\rho_a}  \st^{k_1}_{\sigma_1}  \st^{p_a}_{\lambda_a} \right]
    +{eQ_e\over 2k_1p_b}\; \sum_\rho
           V\left[ \st^{p_b}_{\lambda_b} \st^{k_1}_{\sigma_1} \st^{p_b}_{\rho_b}  \right]
      \Bmf\left[ \st^{p_b}_{\rho_b}  \st^{p_a}_{\lambda_a} \right]\st_{[cd]}\\
   +&{eQ_e\over 2k_1p_a}\; \sum_\rho
     \Bmf\left[\st^{p_b}_{\lambda_b}  \st^{k_1}_{\rho}  \right]\st_{[cd]}
          U\left[\st^{k_1}_{\rho}        \st^{k_1}_{\sigma_1}  \st^{p_a}_{\lambda_a} \right]
    -{eQ_e\over 2k_1p_b}\; \sum_\rho
           V\left[ \st^{p_b}_{\lambda_b} \st^{k_1}_{\sigma_1}  \st^{k_1}_{\rho}      \right]
      \Bmf\left[ \st^{k_1}_{\rho}  \st^{p_a}_{\lambda_a}\right]\st_{[cd]}\\
+ &{\cal M}_{WW\gamma}^1+{\cal M}_{WW\gamma}^2+{\cal M}_{WW\gamma}^3.
  \end{split}
\end{equation}
The term ${\cal M}_{WW\gamma}$ corresponds to the
last three lines%
\footnote{%
  The term ${\cal M}_{WW\gamma}^1+{\cal M}_{WW\gamma}^2+{\cal M}_{WW\gamma}^3$
  originates from the $WW\gamma$ vertex
  \begin{displaymath}
    -ie \bigl[  
    g_{\mu\nu}(p-q)_\rho +g_{\nu\rho}(q-r)_\mu +g_{\nu\rho}(r-p)_\nu 
    \bigr]
  \end{displaymath}
  where all momenta are outcoming, and
  indices on outgoing lines are paired with momenta as
  $p^\mu$, $q^\nu$ $r^\rho$;
  ${\cal M}_{WW\gamma}^1$ originates from the term where $g^{\mu\nu}$ 
  connects the $e^-$--$\nu_e$, $e^+$--$\bar \nu_e$ fermion lines. 
  }
of eq.~(\ref{isr-feynman}). 
These contributions are also IR-finite. 
At this stage  we  keep transfers in the $t$-channel propagators,
which depend on the way how the photon is attached to the fermion line.
The summations in the first two terms get eliminated by the diagonality
property of $U$ and $V$  (see also ref.\cite{Jadach:2000ir}) and we obtain
\begin{equation}
  \label{first-order-isr}
  \begin{split}
   \Meu^{1\{I\}}
      \left( \st^{p}_{\lambda} \st^{k_1}_{\sigma_1} \right)
   =& \sfac^{\{I\}}_{\sigma_1}(k_1) 
  \hat  \Bmf\left[\st^{p}_{\lambda} \right]
   +r_{\{I\}} \left(\st^{p}_{\lambda} \st^{k_1}_{\sigma_1} \right),\\
   r_{\{I\}} \left( \st^{p}_{\lambda} \st^{k_1}_{\sigma_1} \right)=&    
r_{\{I\}}^A + \bigl(  r_{\{I\}}^B
+ {\cal M}_{WW\gamma}^1 \bigr) + \bigl({\cal M}_{WW\gamma}^2+{\cal M}_{WW\gamma}^3\bigr)\\
   r_{\{I\}}^A \left( \st^{p}_{\lambda} \st^{k_1}_{\sigma_1} \right)=&
    +{eQ_e\over 2k_1p_a}\; \sum_\rho
     \Bmf\left[ \st^{p_b}_{\lambda_b} \st^{k_1}_{\rho}   \right]\st_{[cd]}
        U\left[   \st^{k_1}_{\rho} \st^{k_1}_{\sigma_1}  \st^{p_a}_{\lambda_a}   \right]
    -{eQ_e\over 2k_1p_b}\; \sum_\rho
        V\left[   \st^{p_b}_{\lambda_b} \st^{k_1}_{\sigma_1} \st^{k_1}_{\rho}   \right]
     \Bmf\left[ \st^{k_1}_{\rho}    \st^{p_a}_{\lambda_a} \right]\st_{[cd]},\\
   r_{\{I\}}^B \left( \st^{p}_{\lambda} \st^{k_1}_{\sigma_1} \right)= 
   &-{eQ_e\over 2k_1p_a}\; \sum_\rho
     \bar{\Bmf}\left[ \st^{p_b}_{\lambda_b}  \st^{p_a}_{\rho_a}\right]\st_{[cd]}
          U\left[ \st^{p_a}_{\rho_a}  \st^{k_1}_{\sigma_1}  \st^{p_a}_{\lambda_a} \right]
    +{eQ_e\over 2k_1p_b}\; \sum_\rho
           V\left[ \st^{p_b}_{\lambda_b} \st^{k_1}_{\sigma_1} \st^{p_b}_{\rho_b}  \right]
      \bar{\Bmf}\left[ \st^{p_b}_{\rho_b}  \st^{p_a}_{\lambda_a} \right]\st_{[cd]}\\
   \sfac^{\{I\}}_{\sigma_1}(k_1) = &
     -eQ_e{b_{\sigma_1}(k_1,p_a) \over 2k_1p_a} +eQ_e{b_{\sigma_1}(k_1,p_b) \over 2k_1p_b}\quad.
    \end{split}
\end{equation}
The soft part is now clearly separated from the remaining non-IR part,
necessary for the CEEX. 
In $  \hat \Bmf\left[\st^{p}_{\lambda} \right]$ we
use an auxiliary fixed transfer $t_0$, independent of the place where
the photon is attached to the fermion line.
In $\bar{\Bmf}$ we provide the  residual  contribution calculated as 
a difference of the expression calculated with 
the true $t$-transfers $t_a$, $t_b$ and the auxiliary one $t_0$, common 
to all parts of the amplitude.
Note that ${\Bmf} = \hat{\Bmf} + \bar{\Bmf}$.

\subsection{One- and two-loop and one-loop--one-real-photon QED corrections}
In both the one- and two-loop virtual
corrections  we use the same formulae as for the $s$-channel.
This is a very convenient approximation, 
because it simply requires the $W$-exchange amplitudes to be multipied
by already known functions.
As a consequence, the YFS form factor as well as the  $W$ contributions
to the $\beta_0^{(1,2)}$ functions are readily available within the \KK MC environment.

The one-loop, complete, virtual electroweak $W$-exchange contributions
to the $\beta_0^{(1,2)}$ functions are known. 
They are given by the difference between the
exact contribution and the one for the $s$-channel, which is given in 
eq.~(\ref{delta_CC-NC}).
The above virtual $W$-exchange one-loop contribution to $\beta_0^{(1,2)}$
does not include any numerically sizeable terms with respect to 
the scale of the term defined in eq.~(\ref{delta_CC-NC}).
In the program, the one-loop $W$-exchange contribution to $\beta_0^{(1,2)}$
is located in what we call the electroweak $W$ form factor.

Encouraged by the smallness of the above $W$-exchange
one-loop virtual contribution to $\beta_0^{(1,2)}$,
and inspired by the contact interaction approximation,
we assume that the same is true for 
two-loop- and one-loop--one-real-photon QED corrections
in the complete $\beta_0^{(1,2)}$.
At this stage we do not have the complete calculation for $\beta_0^{(1,2)}$.
We assume that the approximate $\beta_0^{(1,2)}$ discussed above
and used in this work differ from the complete ones by
numerically small \Order{\alpha^2} terms; see also~\cite{Berends:1988zz}.
We will thus use the same $s$-channel one-loop contribution to $\beta_0^{(1,2)}$
for the $W$-exchange amplitudes.
This also makes sense because important leading-logarithmic photonic corrections
are universal, i.e. the same for any hard process.

The above approximation is,
of course, our main source of theoretical uncertainty, which we estimate 
to be of the order of 1\% of the cross-sections of the single-photon observables.
As a guide in estimating the size of the above uncertainty 
for a given ${\bar \nu} \nu (n \gamma)$ final state, we use conservatively
the entire size of the one-loop term defined in eq.~(\ref{delta_CC-NC}).

\subsection{Double-photon matrix elements}
Complete double-bremsstrahlung spin amplitudes are at present included. 
Their contribution, as we can see later, turns out to be rather small. 
We will discuss these exact two-photon amplitudes
elsewhere, as well as the related questions of the numerical stability;
see the similar
discussion for the single-bremsstrahlung amplitude described in the next section.
At present, the extensive tests of the type we performed for the $s$-channel amplitudes,
e.g. with the calculation based on  ref.~\cite{Richter-Was:1994ta},
are not yet completed. 
Gauge invariance was used as a main test, so far.
We have noticed numerical stability problems, we could not use the complete amplitudes 
in cases when the photon transverse momenta were smaller than some fraction of the 
electron mass; however, we have checked (for a statistics of 800,000 events)
that this effect is of no numerical relevance.

\subsection{Implementation of photon emission for $W$-exchange in \KK\ MC}
Let us now inspect formula (\ref{first-order-isr}) 
(see also fig.~\ref{fig:bremI}). 
It can be divided into three separately gauge-invariant parts.
The first gauge invariant part is formed out of
the contributions from diagrams (1) and (2) and diagrams (3) and (4)
with the common effective transfer%
\footnote{ In the program, this part of the amplitude
  is calculated  in subroutine {\tt GPS$\_$HiniPlus} and in 
  {\tt GPS$\_$HiniPlusW};
  variables  {\tt Csum1, Csum2}. 
  The contributions $  \bigl(  r_{\{I\}}^B
  + {\cal M}_{WW\gamma}^1 \bigr)$ and  
  $  \bigl({\cal M}_{WW\gamma}^2+{\cal M}_{WW\gamma}^3\bigr)$,
  defined later in the text, are calculated as variables {\tt Csum3} 
  and {\tt Csum4} of subroutine {\tt GPS$\_$HiniPlusW}.} $t_0$:
$\sfac^{\{I\}}_{\sigma_1}(k_1) \Bmf\left[\st^{p}_{\lambda} \right] +r_{\{I\}}^A$.
The second gauge-invariant part includes
the contributions from diagrams (3) and (4), 
responsible for the restoration of true $t_a$ and $t_b$ transfers in place of $t_0$,
combined with part of diagram (5).
It also includes the expression $r_{\{I\}}^B+ {\cal M}_{WW\gamma}^1$
of eq.~(\ref{first-order-isr}).
The third gauge-invariant part is formed out of
the remaining two expressions ${\cal M}_{WW\gamma}^2+{\cal M}_{WW\gamma}^3$
in eq.~(\ref{first-order-isr}).

The formula of eq.~(\ref{first-order-isr}) requires an additional refinement in cases
when more than one photon are present in the event.
In such a case the contribution $r_{\{I\}}^B + {\cal M}_{WW\gamma}^1$ 
would lose gauge invariance because of a ``non conservation'' of the four momenta%
\footnote{%
  For pure $s$-channel amplitudes this complication is absent;
  the propagators of the internal vector bosons do not depend on the individual photon momenta,
  rather on the momenta of the external fermions only,
  see ref.~\cite{Jadach:2000ir} for details.}
$p_a+p_b \ne p_c+p_d + k_1$.
This is cured as follows:
if the momentum carried by other photons is in the same hemisphere as  $p_a$,
we choose
$t_a=(p_b-k_1-p_d)^2$, $t_b=(p_b-p_d)^2$ and
$\epsilon_t=2{\epsilon}^{\mu,\star}_{\sigma_1}(k_1) \cdot ( p_b- p_d)_\mu$;
otherwise we assign
$t_a=(p_a-p_c)^2$, $t_b=(p_a-k_1-p_c)^2$ and
$\epsilon_t=2{\epsilon}^{\mu,\star}_{\sigma_1}(k_1) \cdot ( p_c- p_a)_\mu.$
In addition, the expression ${\cal M}_{WW\gamma}^1 $ is modified as follows:
\begin{equation}
  \label{first-order-isr-reduction}
  {\cal M}^1\left( \st^{p}_{\lambda} \st^{k_1}_{\sigma_1} \right)=\; +e \;\; \;
  \hat{\Bmf}\left[ \st^{p_b}_{\lambda_b}  \st^{p_a}_{\rho_a}\right]          
{\epsilon_t} { {t_0 -M_W^2}\over (t_a -M_W^2)(t_b -M_W^2)},
\end{equation}
which coincides with the original expression, if additional photons are absent.
Also, if the additional photons are collinear with beams, then this choice is
consistent with including them into an ``effective beam'',
in agreement with the principles of the leading-logarithmic approximation.

\subsection{Additional pair correction}
The effects from the emission of real charged pairs accompanying the
$\nu \bar \nu \gamma$  final states may be especially important,
because they may change cross sections and distributions through the non-trivial
experimental event selection (cuts).
It should be kept in mind that
the typical experimental event selection for the
neutrino final state requires that there be no charged track in the detector (veto).
Virtual corrections due to fermion loops in the vertex functions
are included in the \KK\ MC, and
can be switched on as explained in the program documentation. 
The corresponding  real pair emission then needs to be added also.
This can be done by means of a
separate Monte Carlo generation using any massive four-fermion MC event generator,
for example KORALW~\cite{Jadach:2001mp}.

\subsection{Numerical tests using semi-analytical calculations}
Once all necessary ingredients of the  \KK\ Monte Carlo program are explained,
let us start numerical studies, first for the inclusive quantities,
that is for the integrated cross sections.
In table~\ref{KKsem} we show comparisons of results from 
three programs: \KK\ MC, \KK~sem and {\tt ZFITTER}.
We see good agreement in all three final states, $\nu_{\mu,\tau}$ and $\mu$. 
Only in the case of $\nu_e$ ($v_{max}=0.90; 0.99$) was the agreement less satisfactory.
These comparisons provide an important technical test of our program,
although these observables are, of course, academic; their definition 
requires cuts and tagging of invisible neutrinos.

\begin{table}[!h]
\begin{center}
 \caption{The comparison of three programs: \KK MC, \KK\ sem, and {\tt ZFITTER}.}
 \label{KKsem}
\end{center}

\begin{picture}(160,320)
\put( -95, -270){\makebox(0,0)[lb]{\epsfig{file=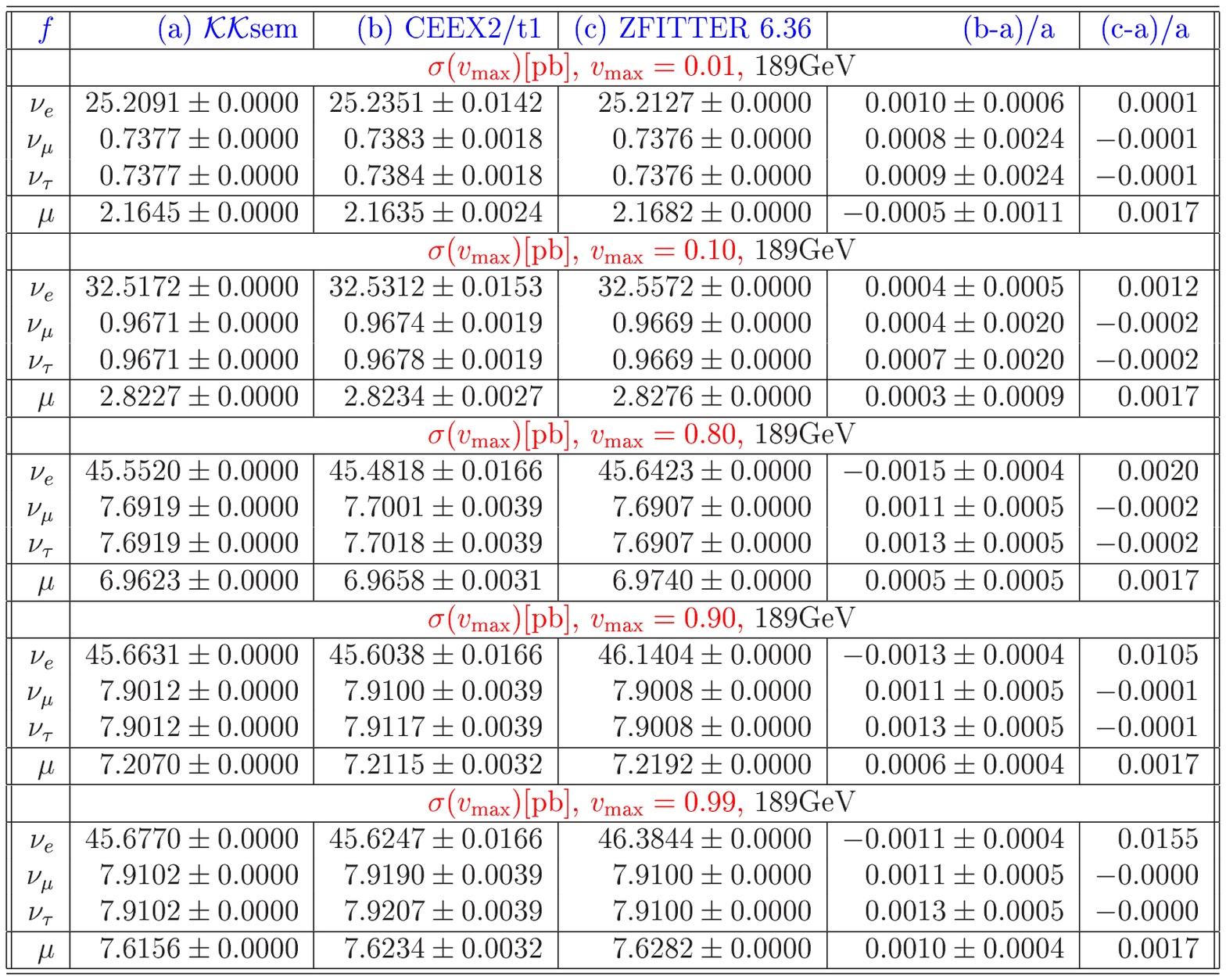,width=190mm,height=250mm}}}
\end{picture} 
\end{table}

\section{Numerical results from \KK\ MC and KORALZ}
Let us now present new important numerical results using observables (event selections)
defined in the LEP2 MC Workshop~\cite{Kobel:2000aw}, without any modifications. 
These observables were defined to suit the needs of all LEP experiments and to match 
their particular experimental studies.

Observables can be divided into two groups:
single photon tagged observables
{\bf Nu1, Nu11, Nu12, Nu13, Nu3, Nu4g, Nu7},
where the precision target required by experiments is  0.5\%, 
and the observables with two observed photons {\bf Nu2, Nu14, Nu5, Nu6, Nu8, Nu9, Nu10},
where the precision target is 2\%, see ref.~\cite{Kobel:2000aw}.

Numerical results are collected in Table 3. 
For every observable and a total energy of 189~GeV, numerical results are collected 
from the \KK\ MC with the exact double hard photon matrix element (CEEX2)
and the exact single matrix element only (CEEX1).
First, we can conclude that the predictions from 
KORALZ presented during the LEP2 MC Workshop and for all neutrino observables
were generally correct within 4\%, in agreement with what has been declared. 
One can even get an impression that the agreement is better, of the order of 2\%. 
This, however, is misleading, since 
these small differences were due to accidental cancellations of several effects,
all of them of the order of 2--3\% .

Let us now discus results from  the \KK\ MC with the various
options for the matrix element, having in mind an estimate of the theoretical uncertainties. 
We have found for all observables that
the correction from additional virtual fermion pairs was systematically $-0.6\%$.
(The corresponding entries are not included in the actual version of table 3.)
Since the virtual fermion pairs contribution is rather small, in comparison
with other uncertainties, we will use the size of the above virtual
correction of $0.6\%$ as a conservative estimate of the corresponding systematic error. 
Concerning the missing higher order QED  contributions
(see section 3.4) we estimate it to be 1\% for $\nu_e$ and 0.5\% for other neutrinos.
Finally we take 0.4\% as for the systematic uncertainty of electroweak corrections of ZFITTER. 
The final uncertainty for the single-photon observables is thus estimated to be 1.3\% for 
electron neutrino final states and about 0.8\% for the other neutrino species.

The uncertainty for observables with two tagged hard photons is higher, 
probably of the order of%
\footnote{%
  Such an estimate was obtained by switching on and off 
  the double-photon contribution to the matrix element. See entries CEEX2 and CEEX1 in Table 3.  
  Also, the different options of reduction procedures could be compared
  and the corresponding contribution to the net uncertainties could be reduced. 
  We leave, however, these points for future studies.}
5\%.
As a numerical input we use standard initialization parameters of the \KK\ MC defined in 
ref.~\cite{Jadach:1999vf}. 


{\small
\begin{table}[!ht]
\small
\centering
\caption{\small
Numerical predictions for observables of the LEP2 MC Workshop
at a total energy of 189~GeV (input data of the Workshop).
For each observable results are shown in 3 lines:
(i) CEEX2 ${\cal O}(\alpha^2)$ \KK\ MC,
(ii) CEEX1 ${\cal O}(\alpha^1)$ \KK\ MC,
(iii) KORALZ MC.
The difference $\delta$ is shown
in the last column: it is the deviation from 1 ($\times 100$)    
of the ratio of a given result to that of CEEX2.                                    
}
\begin{tabular} {||l|l|l|l||}
\hline\hline
Label of obs.                      &
Program                            &
Cross section [pb]                 &
$\delta$ ratio                     
\\
\hline
\hline
 \citobs{Nu1}   
  & KKMC     CEEX2 & 
 3.1710$\cdot 10^{ 0\;\; }\ \pm$7.43$\cdot 10^{-3     }$
 &  
\\
 
  & KKMC     CEEX1 & 
 3.2139$\cdot 10^{ 0\;\; }\ \pm$3.85$\cdot 10^{-3     }$
 &   1.35
\\
 
  & KORALZ 4.04    & 
 3.2244$\cdot 10^{ 0\;\; }\ \pm$4.34$\cdot 10^{-3     }$
 &   1.68
\\
\hline
 \citobs{Nu2}   
  & KKMC     CEEX2 & 
 2.1886$\cdot 10^{-1     }\ \pm$9.26$\cdot 10^{-4     }$
 &  
\\
 
  & KKMC     CEEX1 & 
 2.1655$\cdot 10^{-1     }\ \pm$8.79$\cdot 10^{-4     }$
 &  -1.06
\\
 
  & KORALZ 4.04    & 
 2.1733$\cdot 10^{-1     }\ \pm$1.20$\cdot 10^{-3     }$
 &  -0.70
\\
\hline
 \citobs{Nu11}   
  & KKMC     CEEX2 & 
 9.1551$\cdot 10^{-1     }\ \pm$2.26$\cdot 10^{-3     }$
 &  
\\
 
  & KKMC     CEEX1 & 
 9.1066$\cdot 10^{-1     }\ \pm$2.23$\cdot 10^{-3     }$
 &  -0.53
\\
 
  & KORALZ 4.04    & 
 9.1767$\cdot 10^{-1     }\ \pm$2.42$\cdot 10^{-3     }$
 &   0.24
\\
\hline
 \citobs{Nu12}   
  & KKMC     CEEX2 & 
 1.8242$\cdot 10^{ 0\;\; }\ \pm$3.09$\cdot 10^{-3     }$
 &  
\\
 
  & KKMC     CEEX1 & 
 1.8359$\cdot 10^{ 0\;\; }\ \pm$3.07$\cdot 10^{-3     }$
 &   0.64
\\
 
  & KORALZ 4.04    & 
 1.8442$\cdot 10^{ 0\;\; }\ \pm$3.37$\cdot 10^{-3     }$
 &   1.09
\\
\hline
 \citobs{Nu13}   
  & KKMC     CEEX2 & 
 1.7775$\cdot 10^{ 0\;\; }\ \pm$6.95$\cdot 10^{-3     }$
 &  
\\
 
  & KKMC     CEEX1 & 
 1.8085$\cdot 10^{ 0\;\; }\ \pm$2.82$\cdot 10^{-3     }$
 &   1.75
\\
 
  & KORALZ 4.04    & 
 1.8045$\cdot 10^{ 0\;\; }\ \pm$3.35$\cdot 10^{-3     }$
 &   1.52
\\
\hline
 \citobs{Nu14}   
  & KKMC     CEEX2 & 
 2.0634$\cdot 10^{-1     }\ \pm$9.02$\cdot 10^{-4     }$
 &  
\\
 
  & KKMC     CEEX1 & 
 2.0267$\cdot 10^{-1     }\ \pm$8.40$\cdot 10^{-4     }$
 &  -1.78
\\
 
  & KORALZ 4.04    & 
 2.0121$\cdot 10^{-1     }\ \pm$1.15$\cdot 10^{-3     }$
 &  -2.49
\\
\hline
 \citobs{Nu3}   
  & KKMC     CEEX2 & 
 4.2434$\cdot 10^{ 0\;\; }\ \pm$7.78$\cdot 10^{-3     }$
 &  
\\
 
  & KKMC     CEEX1 & 
 4.2912$\cdot 10^{ 0\;\; }\ \pm$4.48$\cdot 10^{-3     }$
 &   1.13
\\
 
  & KORALZ 4.04    & 
 4.2885$\cdot 10^{ 0\;\; }\ \pm$4.89$\cdot 10^{-3     }$
 &   1.06
\\
\hline
 \citobs{Nu5}   
  & KKMC     CEEX2 & 
 1.2128$\cdot 10^{-1     }\ \pm$6.77$\cdot 10^{-4     }$
 &  
\\
 
  & KKMC     CEEX1 & 
 1.1912$\cdot 10^{-1     }\ \pm$6.26$\cdot 10^{-4     }$
 &  -1.78
\\
 
  & KORALZ 4.04    & 
 1.1850$\cdot 10^{-1     }\ \pm$8.83$\cdot 10^{-4     }$
 &  -2.30
\\
\hline
 \citobs{Nu6}   
  & KKMC     CEEX2 & 
 5.7331$\cdot 10^{-2     }\ \pm$4.64$\cdot 10^{-4     }$
 &  
\\
 
  & KKMC     CEEX1 & 
 5.5817$\cdot 10^{-2     }\ \pm$4.22$\cdot 10^{-4     }$
 &  -2.64
\\
 
  & KORALZ 4.04    & 
 5.6543$\cdot 10^{-2     }\ \pm$6.11$\cdot 10^{-4     }$
 &  -1.37
\\
\hline
 \citobs{Nu7}   
  & KKMC     CEEX2 & 
 4.4676$\cdot 10^{ 0\;\; }\ \pm$7.82$\cdot 10^{-3     }$
 &  
\\
 
  & KKMC     CEEX1 & 
 4.5206$\cdot 10^{ 0\;\; }\ \pm$4.54$\cdot 10^{-3     }$
 &   1.19
\\
 
  & KORALZ 4.04    & 
 4.5109$\cdot 10^{ 0\;\; }\ \pm$5.00$\cdot 10^{-3     }$
 &   0.97
\\
\hline
 \citobs{Nu8}   
  & KKMC     CEEX2 & 
 1.7593$\cdot 10^{-1     }\ \pm$8.31$\cdot 10^{-4     }$
 &  
\\
 
  & KKMC     CEEX1 & 
 1.7162$\cdot 10^{-1     }\ \pm$7.63$\cdot 10^{-4     }$
 &  -2.45
\\
 
  & KORALZ 4.04    & 
 1.7074$\cdot 10^{-1     }\ \pm$1.06$\cdot 10^{-3     }$
 &  -2.95
\\
\hline
 \citobs{Nu9}   
  & KKMC     CEEX2 & 
 7.6434$\cdot 10^{-2     }\ \pm$5.39$\cdot 10^{-4     }$
 &  
\\
 
  & KKMC     CEEX1 & 
 7.4473$\cdot 10^{-2     }\ \pm$4.89$\cdot 10^{-4     }$
 &  -2.56
\\
 
  & KORALZ 4.04    & 
 7.4208$\cdot 10^{-2     }\ \pm$7.00$\cdot 10^{-4     }$
 &  -2.91
\\
\hline
 \citobs{Nu10}   
  & KKMC     CEEX2 & 
 2.5362$\cdot 10^{-1     }\ \pm$1.00$\cdot 10^{-3     }$
 &  
\\
 
  & KKMC     CEEX1 & 
 2.5107$\cdot 10^{-1     }\ \pm$9.39$\cdot 10^{-4     }$
 &  -1.01
\\
 
  & KORALZ 4.04    & 
 2.5040$\cdot 10^{-1     }\ \pm$1.28$\cdot 10^{-3     }$
 &  -1.27
\\
\hline
 \citobs{Nu4g}   
  & KKMC     CEEX2 & 
 1.9091$\cdot 10^{ 0\;\; }\ \pm$6.99$\cdot 10^{-3     }$
 &  
\\
 
  & KKMC     CEEX1 & 
 1.9379$\cdot 10^{ 0\;\; }\ \pm$2.92$\cdot 10^{-3     }$
 &   1.51
\\
 
  & KORALZ 4.04    & 
 1.9398$\cdot 10^{ 0\;\; }\ \pm$3.46$\cdot 10^{-3     }$
 &   1.61
\\
\hline\hline
\end{tabular}
\end{table}
}

\section{Conclusions}
We have extended the Monte Carlo program \KK\ to the neutrino mode.
The systematic error is estimated to be
1.3\%  for $\nu_e \bar \nu_e \gamma$ and 0.8\% for 
$\nu_\mu \bar \nu_\mu \gamma $ and $ \nu_\tau\bar \nu_\tau \gamma$.
For observables with two observed photons we estimate the uncertainty to
be about 5\%.
These new improved results were obtained thanks to
the inclusion of non-photonic electroweak corrections
of the {\tt ZFITTER} package~\cite{Bardin:1999yd,home-ZFITTER}
and due to newly constructed, exact,
single and double emission photon amplitudes in the \KK\ MC
for the contribution with the $t$-channel $W$ exchange.
The virtual corrections for
the $W$ exchange are at present introduced in the approximated form.
The exponentiation scheme CEEX is the same as in the original \KK\ MC program
of refs.~\cite{Jadach:1999vf,Jadach:1998jb,Jadach:2000ir}.

Let us also note that in the other Monte Carlo programs
available in the literature for the neutrino channel,
see eg. refs.~\cite{Kurihara:1999vc,Montagna:2001ej},
the exact double-photon bremsstrahlung amplitudes are also included. 
Direct comparisons with these programs should be done at a certain point.
This could provide an independent cross check of our
double-photon matrix element and may also shed light on certain
non-negligible differences between the results for the neutrino channel
LEP observables collected from various MC event generators
during the LEP2 MC Workshop~\cite{Kobel:2000aw}.
Therefore, a better answer could be provided on the total theoretical
and technical uncertainties in these calculations.

\section*{Acknowledgements}

Z.W. would like to thank Paul Colas for his encouraging to start studying the 
phenomenology of $e^+e^- \to \bar \nu \nu \gamma$ in the early 90's. 
Useful discussions with B.~Bloch, J. Boucrot, A.~Jacholkowska, W.~P\l{}aczek, M. Skrzypek and 
B.F.L.~Ward are also acknowledged.

\bibliographystyle{utphys_spires}

\providecommand{\href}[2]{#2}
\end{document}